\documentstyle[12pt,epsf,epsfig]{article}
\begin{document}


\begin{center}
\Large
\vskip 2truecm

Towards Low-Threshold, Real-Time Solar Neutrino Detectors

\vskip 2truecm
\large

G. Gratta, Y.F. Wang 

\vskip 0.5truecm

 Physics Department, Stanford University

\begin{abstract}

We discuss an alternative approach to the detection of solar neutrinos
using a coarsely segmented detector based on inverse-beta decay onto $^{160}$Gd or $^{176}$Yb.
While it is know that similar approaches, already discussed in the literature, 
can in principle provide low-threshold, real-time energy spectroscopy with intrinsic background
rejection features, the concepts presented here make this scheme possible with lower
background and current technology. 

\end{abstract}

\normalsize

\end{center}

\vfill
\eject

\normalsize

Neutrinos from the sun have been observed by a number of detectors
over the past 30 years. While it is generally accepted that the 
Standard Solar Model(SSM)~\cite{Bahcall} has been remarkably successful in 
describing the nuclear fusion in the sun, large discrepancies
remain between experimental data and the SSM predictions~\cite{deficit}.
Indeed it is quite possible that the solar neutrino anomaly be due to
phenomena related to intrinsic neutrino properties.
Detailed understanding of solar neutrinos is limited by our ability to 
collect high quality data on the subject, so that further progress in the 
field requires the development of qualitatively new experimental tools.

According to the SSM the solar neutrino flux is dominated by the 
low energy (0-0.4 MeV) pp component which has been measured by 
the Gallex and SAGE experiments. These two experiments, however, can only 
measure the integrated flux above an energy of 0.23 MeV.   Similarly the
pioneering chlorine experiment at Homestake integrates all energies above 0.81 MeV,
hence including most of $^7$Be along with pep and $^8$B fluxes.
Water \v{C}erenkov detectors like (Super)Kamiokande~\cite{SK} and SNO~\cite{SNO}, 
and large scintillation detectors like Borexino~\cite{borexino} and 
KamLAND~\cite{kamland} can do real time spectroscopy but with a resolution limited
by the electron scattering kinematics.   More important their thresholds are  
too high to be sensitive to pp neutrinos.

In order to make a qualitative improvement in the field one has to find
a detection scheme able to provide real-time detection with low threshold to
include the pp flux and good energy resolution.     The main obstacle in achieving
these goals comes from the formidable backgrounds from natural radioactivity
that severely limit the energy threshold of scintillation and \v{C}erenkov detectors.
Such backgrounds force one to adopt extremely large shielding volumes and
levels of U, Th, $^{40}$K and $^{14}$C in the active scintillation fluid 
that are at the very limit of what is measurable\cite{borexino, kamland}.
While Borexino has demonstrated their target contamination levels 
($10^{-16}$ g/g for U/Th, $10^{-18}$ g/g for $\mathrm ^{40}K$
and $10^{-18}$ for $^{14}$C/$^{12}$C) in a small size test facility~\cite{ctf}
the possibility of achieving similar values in a few-hundred-times larger detector
remains to be demonstrated.    Indeed it is quite clear that backgrounds
(in particular from $^{14}$C) grow too rapidly below 400 keV to try to perform
any measurement below the pp threshold.

A new detection scheme able to fulfill all the requirements mentioned above
has been recently proposed~\cite{raghavan}.  In such a scheme $^{160}$Gd and
$^{176}$Yb are used as targets for solar neutrino absorption through the
inverse-$\beta$ decay reactions
$$\mathrm \nu_e + ^{160}Gd (^{176}Yb) \longrightarrow e^- + ^{160}Tb^* (^{176}Lu^*).$$
The $\mathrm ^{160}Tb^* (^{176}Lu^*)$ states de-excite with a lifetime of 87 ns (50 ns)
emitting $\gamma$s of 63.7 keV (72 keV). 
The threshold of these reactions is 0.24 MeV (0.30 MeV), well below the pp 
threshold, and while the {\it total} neutrino energy can be measured from the 
energy release by electron and $\gamma$ in the inverse-$\beta$ process, the 
fast correlation time between electron and $\gamma$ emission provides a powerful
method for suppressing backgrounds.
The target isotopes would be dissolved in liquid scintillator to provide 
high detection efficiency and good energy resolution for the low energy processes described.
While precise measurements of the neutrino capture cross sections for the 
above reactions will be essential for a successful experiment, it is
suggested\cite{ftvalues} that the $\log{ft}$ are in the range 3.5 - 4.2
so that a 100 ton scintillation detector, loaded with 10\% of natural Gd or Yb
would give a signal of the order of 1 event/day.
As suggested in~\cite{raghavan} spatial segmentation of the detector provides
an additional suppression of the background since electron and $\gamma$ emission
in the neutrino capture processes are bound to originate from the same region of space.

In this letter we will point out some difficulties with this scheme and suggest
ways to bring this powerful idea to full fruition.

The addition of Gd (Yb) to the scintillator poses two main problems:
\begin{itemize}
  \item while several chemical processes able to load liquid scintillators
        with rare-earths and other heavy elements are know\cite{paloverde,
        chooz,bicron,raghavan} we believe that it is quite unlikely that at the high
        loading required (10\%-30\%) the scintillator can still
        be transparent and stable enough to
        perform a long term experiment with a detector of the size required.
        The last generation of reactor neutrino oscillation experiments, and
        in particular the Palo Verde detector, are very similar in size and 
        need to detect very low energy events like in the proposed scheme.   Both
        Chooz~\cite{chooz} and Palo Verde~\cite{paloverde} 
        used 0.1\% Gd loading and, while attenuation
        lengths of $\simeq 10$m and stable operation were finally obtained, it is
        unlikely that an increase in loading of few hundred times
        could still give a scintillator capable of detecting a fraction of 100 keV
        at several meter distance for many years.
  \item it is known that the purification of rare 
        earths from heavy metals such as Th and U is relatively more difficult
        than in the case of the organic components of scintillators.
        Hence a highly loaded scintillator is bound to be substantially ``dirtier''
        than the organic components alone.
        Some measurements~\cite{andreas} of high quality commercial
        Gd oxides and nitrates give U and Th contaminations at the level of 
        0.8 to 20~ppb.   If we aggressively assume a factor of ten improvement
        on the best figure (0.1~ppb) we still obtain a $10^{-11}$ g/g contamination 
        for the whole liquid scintillator (at 10\% loading). 
        While the time and spatial correlations provide uncorrelated background
        suppression factors of, respectively, $10^{8}$ and $10^{4}$, such reductions
        do not apply to decay sequences from natural isotopes having
        lifetimes in the 10-1000~ns range.   Unfortunately one such cases is present 
        in the $^{232}$Th decay chain:
        $$\mathrm ^{212}Bi\longrightarrow ^{212}Po(299 ~ns)\longrightarrow ^{208}Pb$$
        While the Bi  $\beta$-decays with a $Q$ value of about 2.5 MeV the Po decays
        with a 8.8 MeV $\alpha$.   
        We find that substantial
        background arises from the cases in which the Po decay occurs near the walls
        of the container so that part of the ionization from the $\alpha$ is lost 
        into passive material.     We note here that, while the $\alpha$ energy is
        well above the energy expected for neutrino events, scintillation light
        is generally quenched by a large factor of order 10.
        Pulse shape analysis may improve slightly the situation but will certainly be 
        hampered by the non-optimal light collection geometry.  Our detailed simulation
        predicts a rate of mis-identified events of 10~day$^{-1}$ for a cell cross-section
        20 $\times$ 20 cm$^2$ (and the mass and contamination figures indicated above).
        It is interesting to note that this background derives {\it only} from the 
        outermost layer of scintillator (about 100~$\mu$m thick) that can give 
        non-contained $\alpha$s. Hence, while finer segmentation improves the rejection
        of the uncorrelated backgrounds, at the same time it increases the surface to volume 
        ratio making the effects of the Bi-Po background worse.
\end{itemize}

Both difficulties would be completely eliminated by using
transparent and active material, such as plastic scintillator, as the walls of liquid 
scintillator cells.   The plastic scintillator would serve two distinct purposes:
\begin{itemize}
\item[1)] it would fully contain the $\alpha$s produced by the impurities in the Gd, 
\item[2)] it would act as a wavelength shifter and absorb, re-emit and transport the 
          light produced by scintillation in the Gd-doped scintillator. 
\end{itemize}
\begin{figure}[hbt]
\begin{center}
\mbox{\epsfig{file=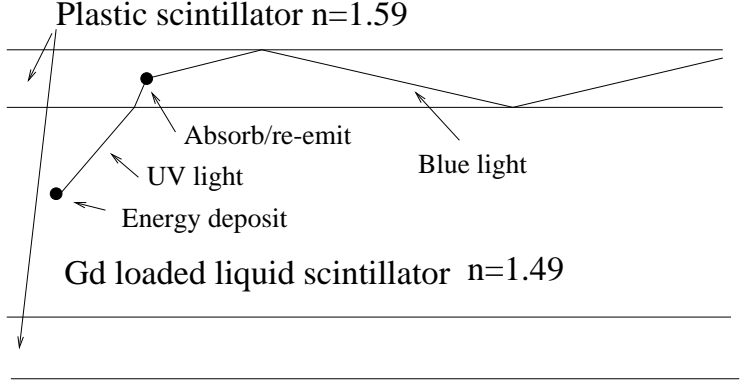,height=5.cm}}
\caption{The basic principle of the detector (dimensions not to scale).}
\end{center}
\end{figure}
Fig. 1 shows the basic principle of the detector.
Plastic wavelength shifters have been used in a similar fashion in a number of
cases\cite{shifters} mainly in order to reduce the active surface (and hence, in 
general, the noise) of the light detector in scintillation counters.    In our scheme
an appropriate fluor in the plastic scintillator would act as a shifter for the light
produced with shorter wavelength in the liquid scintillator.    The longer wavelength light
will then be transported inside the plastic allowing us to tolerate 
loaded scintillators with light attenuation length of the order of the transverse cell 
size (i.e. $\sim$ 50 cm).   Plastic scintillators with 5 m light attenuation length are
common and commercially available~\cite{bicron}.
The use of a complete plastic scintillator (as opposed to a simple shifter) essentially
provides a fully active medium with cleaner conditions at the surface where $\alpha$s
are difficult to reject.   Photo-detectors (assumed to be conventional photomultipliers in our
analysis) should be fitted onto the ends of the plastic scintillator walls only.
A further potential advantage, that we do not study in detail here, may lay in better
pulse-shape discrimination obtained by guiding the light through a smaller cross-section
channel (the plastic scintillator wall as opposed to the bulk liquid).

In order for this system to work properly some substantial fraction of light has to be 
extracted from the liquid into the plastic and then be trapped in there.   This can be 
achieved by a proper match of refractive indices, with $n_{liquid} < n_{plastic}$ 
(and, trivially,  $n_{air} < n_{plastic}$).   Although a careful R\&D study may provide 
better figures here we assume $n_{liquid} = 1.49$ and $n_{plastic} = 1.59$ as it is the
case for standard commercial products.   We perform a full Monte Carlo simulation using 
GEANT3 to study in detail the signal detection and the resolution and to estimate the 
background and the light collection efficiency. Each detector cell is 
$\mathrm 20 \times 20 \times 500 cm^3$, a total of 625 cells constitute the detector 
with total dimensions of $\mathrm 5 \times 5\times 5~m^3$ and a total mass of 125~tons. 
The plastic scintillator walls are assumed to be 1~cm thick.
Furthermore we assume the yield of the loaded scintillator to be  
5000 photons per MeV (roughly 50\% of standard liquid scintillators) and the 
conversion quantum efficiency of the scintillating walls 80\%.
These factors, together with the optical efficiency of the system and quantum efficiency of
a bi-alkali photocathode, give 1.8 PE/50 keV at each end of a detector cell. This is
sufficient to make a coincidence of the signals from the two ends to reduce the photo-detector 
noise. 
Fig.~2 shows the simulated energy spectrum of $\nu_e$ as would result  from this system.
Both pp and $^7$Be peaks are visible.   We note here that Gd gives a substantially better
pp detection due to the lower threshold.
\begin{figure}[hbt]
\begin{center}
\mbox{\epsfig{file=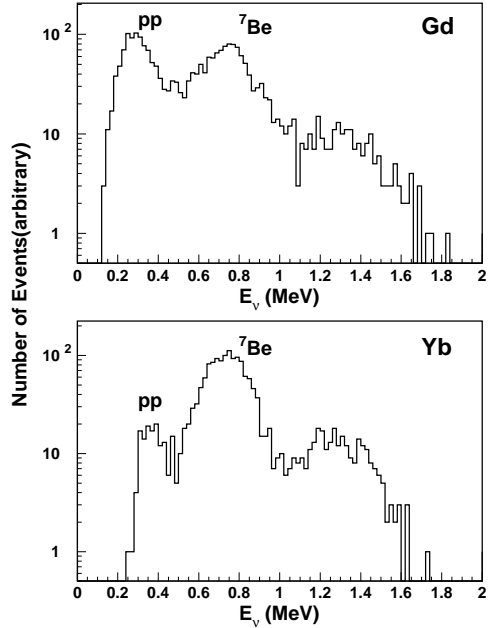,height=10.cm}}
\caption{Simulated energy spectrum of $\nu_e$ from the detector concept presented.
The lower threshold in Gd gives a substantially
better pp-neutrino detection.}
\end{center}
\end{figure}

Table 1 gives our estimate for the uncorrelated background coming from the natural 
radioactivity. While the rate $N_1$ corresponds to energies between 0.04-2~MeV 
which could mimic the prompt $\beta$, $N_2$ corresponds to energies between 
0.04-0.16~MeV which could mimic the delayed low energy $\gamma$. 
Using a coincidence window of 250 ns, and a space reduction factor of 
$10^{-4}$ (derived from the transverse segmentation and a longitudinal resolution 
of $\simeq 50$~cm), we obtain a background rate of 0.1/day, well below the signal rate.
We note here that our table does not include backgrounds arising from
possible $^{176}$Lu and $^{147}$Sm contamination in the Yb and Gd compounds.
Special chemical processing will have to be devised in order to make our scheme
(and probably any other scheme based on a rare-earth target) viable~\cite{raghavan}.

The correlated background from the Bi-Po chain is now negligible provided
that the Th concentration in plastic scintillator is below $10^{-13}$g/g.    This
requirement is readily achievable with commercial materials.
We note here that, while a somewhat similar scheme employing wavelength-shifting
fibers was proposed in~\cite{raghavan_fibers}, our purpose is to efficiently extract
light from a large detector while reducing the background from correlated sources.
This second purpose in fact substantially impacts the detector optimization, 
making a very fine segmentation a losing strategy.

\begin{table}[htbp]
\begin{center}
\begin{tabular}{|c|c|c|c|}
\hline
Isotope & Purity & $N_1$ (Hz) & $N_2$ (Hz) \\  
\hline
$^{238}$U  &  $10^{-11}$ g/g & 150  & 15   \\
$^{232}$Th &  $10^{-11}$ g/g & 31   & 2.6   \\
$^{40}$K   &  $10^{-11}$ g/g & 250  & 12   \\
$^{222}$Rn &  $0.1 $Bq/m$^3$   & 88   & 6   \\
$^{14}$C   &  $10^{-18}$  $^{14}$C/$^{12}$C  & 14  & 14   \\
\hline
Total      &                 &  533 & 50 \\
\hline
\end{tabular}
\end{center}
\end{table}

The chemical compatibility between loaded liquid scintillators and a
plastic scintillator container may appear to be problematic since aromatic solvents 
(that are generally used to achieve high 
loading) are aggressive, in long periods of time, for polystyrene-based plastics.
While it is quite possible that a proper study will be able to produce a pair of compatible
materials we briefly describe here three alternative techniques to avoid the problem 
altogether.
\begin{itemize}
\item Very thin (few~$\mu$m thick) fluoropolymer coatings are available in completely
transparent form (over 95\% transmission in the near UV) as separate films to apply 
to the surface to protect or through a chemical deposition process.   
Such fluoropolymer coatings would chemically isolate 
the liquid scintillator from the plastic still retaining all the features described (optics 
and almost full $\alpha$ detection).   
\item Highly loaded liquid scintillator
cocktails are possible by forming a micro-emulsion of a scintillating fluid (containing heavy
and hence non aggressive aromatics) with a water-based solution of the rare-earth 
element~\cite{packard_trick}.    Such systems, commonly used in the life sciences typically
have poor light attenuation lengths that are, however, perfectly adequate in our case.
Indeed refractive indices as low as 1.44 can be achieved with this technique~\cite{terwiel}.
\item The Gd (Yb) loaded part of the detector could also be built out of
plastic scintillator.    In this case there would probably be a very small air gap
between the loaded core and the unloaded ``box'' around and, although the optics will be
somewhat different, we believe that our conclusions would still hold.
\end{itemize}

In summary we have presented an alternative approach to the detection of solar neutrinos
using a coarsely segmented detector based on inverse-beta decay onto $^{160}$Gd or $^{176}$Yb.
Such approach would allow low-threshold, real-time energy spectroscopy with technology 
available today.    The correlated signature of the neutrino events would allow effective
reduction of random backgrounds, while the remaining correlated backgrounds will be
eliminated by the particular detector configuration.

We would like to thank Drs. J.~Bahcall, F.~Boehm, A.~Piepke, R.~Raghavan and P.~Vogel 
for many interesting discussions on the subject. We particularly thank Dr. R.~Raghavan for 
pointing us to ref.~\cite{raghavan_fibers}.
This work has been supported in part by DoE Grant DE-FG03-96ER40986 and NSF Grant PHY-960-1613.


\begin{thebibliography}{99}

\bibitem{Bahcall}See J.N. Bahcall, ``Neutrino Astrophysics'', Cambridge
University Press, Cambridge, 1990.

\bibitem{deficit} See Particle Data Group, ``Solar $\nu$ Experiments'',
The Europ. Phys. Jour. C {\bf 3}, 327 (1998).


\bibitem{SK} For considerations on the Kamiokande and SuperKamiokande detectors
             see for example: K. Nakamura {\it et al.} p. 249-387
             in ``Physics and Astrophysics
             of Neutrinos, Eds. M. Fukugita, A. Suzuki; Springer-Verlag.
          
\bibitem{SNO} B.C. Robertson {\it et al.} p.119-125, in ``Perspectives in neutrinos, atomic
               physics and gravitation'', Villars sur Ollon 1993.

\bibitem{borexino} M.G. Giammarchi {\it et al.}, p 103-111, in Les Arcs 1992: ``Progress
                   in Atomic Physics, Neutrinos and Gravitation'', Editions Frontieres.

\bibitem{kamland} P. Alivisatos {\it et al.}, ``KamLAND: a Liquid scintillator Anti-Neutrino
Detector at the Kamioka site'', Stanford-HEP-98-03/Tohoku-RCNS-98-15, unpublished.

\bibitem{ctf}G. Ranucci {\it et al.}, Nucl. Phys. Proc. Suppl. {\bf 70}, 377(1999).

\bibitem{raghavan}R.S. Raghavan, Phys. Rev. Lett. {\bf 78} 3618(1997).

\bibitem{ftvalues} R.S. Raghavan, to appear in ``Proceedings of the 18th International Conference
                   on Neutrino Physics and Astrophysics'', Takayama, Japan, June 1998.


\bibitem{paloverde} A. Piepke, V. Novikov and W. Moser, ``Development of a Gd-loaded scintillator
                       for for electron anti-neutrino spectroscopy'', to appear in NIM.

\bibitem{chooz} M. Apollonio {\it et al.} Phys. Lett. B420 (1998) 397.

\bibitem{bicron} Bicron SGC, Organic Scintillators Catalog 1997, p. 8.

\bibitem{andreas} A. Piepke, private communication.

\bibitem{shifters} E. Aker {\it et al.} Nucl. Instr. and Meth. A321 (1992) 69;
                   GEM Technical Design Report, SSCL-SR-1219, unpublished.

\bibitem{raghavan_fibers} R.S. Raghavan in ``Proceedings of the 1981 International 
                          Conference on Neutrino Physics and Astrophysics, 
                          Maui, Hawaii, July 1-8, 1981 (ed. by R.J.~Cence, E.~Ma, A.~Roberts),
                          p.27.



\bibitem{packard_trick} J. Thomson, p. 408-414, in ``Handbook of Radioactivity Analysis'',
                        Ed. M. L'Annunziata, Academic Press 1998;
                        M.S. Patterson and R.C. Green, Analytical Chemistry 37 (1965) 854.

\bibitem{terwiel} J. terWiel, private communication.

\end{thebibliography}
\end{document}